\documentclass[a4paper,man,natbib,floatsintext]{apa6}
\usepackage{natbib}
\usepackage{xcolor}
\usepackage{tabularx}
\usepackage{makecell}
\usepackage{graphicx} 
\usepackage{subcaption}
\usepackage[english]{babel}
\usepackage[autostyle, english = american]{csquotes}
\MakeOuterQuote{"}

\usepackage{verbatim}

\usepackage{caption}
\usepackage[font=footnotesize]{caption}
\usepackage{amsmath}
\usepackage{lipsum} 
\usepackage{hyperref}
\usepackage{multirow}

\usepackage{adjustbox}
\usepackage{longtable}
\usepackage{booktabs}
\usepackage{enumitem}
\setlist[itemize]{leftmargin=*}

\DeclareUnicodeCharacter{2212}{-}

\begin{document}
%

\begin{titlepage}

\begin{center}

\large The Mediating Effects of Emotions on Trust through Risk Perception and System Performance in Automated Driving
\\ 

\normalsize

\vspace{25pt}
Lilit Avetisyan\\
Industrial and Manufacturing Systems Engineering, University of Michigan-Dearborn\\
Emmanuel Abolarin\\
Industrial and Manufacturing Systems Engineering, University of Michigan-Dearborn\\
Vanik Zakarian\\
Industrial and Manufacturing Systems Engineering, University of Michigan-Dearborn\\
\vspace{15pt}
X. Jessie Yang\\
Industrial and Operations Engineering, University of Michigan-Ann Arbor\\
\vspace{15pt}
Feng Zhou\\
Industrial and Manufacturing Systems Engineering, University of Michigan-Dearborn\\
\vspace{15pt}
\end{center}
\begin{flushleft}
\textbf{Manuscript type:} \textit{Research Article}\\
\textbf{Running head:} \textit{ Mediating Effects of Emotions on Trust in  Automated Driving}\\
\textbf{Word count:} 8471 \\ 

\textbf{Corresponding author:} 
Feng Zhou, 4901 Evergreen Road, Dearborn, MI 48128, Email: fezhou@umich.edu


\end{flushleft}

\end{titlepage}
\shorttitle{}



\section{ABSTRACT}
Trust in automated vehicles (AVs) has traditionally been explored through a cognitive lens, but growing evidence highlights the significant role emotions play in shaping trust. This study investigates how risk perception and AV performance (error vs. no error) influence emotional responses and trust in AVs, using mediation analysis to examine the indirect effects of emotions. In this study, 70 participants (42 male, 28 female) watched real-life recorded videos of AVs operating with or without errors, coupled with varying levels of risk information (high, low, or none). They reported their anticipated emotional responses using 19 discrete emotion items, and trust was assessed through dispositional, learned, and situational trust measures. Factor analysis identified four key emotional components, namely hostility, confidence, anxiety, and loneliness, that were influenced by risk perception and AV performance. The linear mixed model showed that risk perception was not a significant predictor of trust, while performance and individual differences were. Mediation analysis revealed that confidence was a strong positive mediator, while hostile and anxious emotions negatively impacted trust. However, lonely emotions did not significantly mediate the relationship between AV performance and trust. The results show that real-time AV behavior is more influential on trust than pre-existing risk perceptions, indicating trust in AVs might be more experience-based than shaped by prior beliefs. Our findings also underscore the importance of fostering positive emotional responses for trust calibration, which has important implications for user experience design in automated driving.


\textbf{Keywords:} Emotion, trust, risk, physiological measures, automated vehicles.

\newpage
\section{Introduction}


The rapid advancement in automated vehicle (AV) technology has made it increasingly evident that Level 3+ AVs will soon become widely available and affordable. Deichmann \citeyearpar{deichmann2023autonomous} projects that 12\% of new passenger cars will feature Level 3+ autonomous technologies by 2030, rising to 37\% by 2035. However, consumer willingness to adopt AVs has decreased, with only 26\% of respondents in 2021 expressing a preference for fully automated vehicles, compared to 35\% in 2020. This decline in adoption intent highlights a critical challenge: establishing and maintaining user trust in AVs. Trust has become a central focus of research in human-AV interaction, particularly as AVs are integrated into everyday life \citep{ayoub2019manual, Ayoub2021Modeling, ayoub2023real, avetisyan2024building}. While AVs promise benefits such as increased safety and comfort, inappropriate levels of trust (either overtrust or undertrust) can lead to misuse of unreliable systems or disuse of reliable ones.

Automation trust is a fundamental psychological mechanism governing human engagement with automated systems \citep{kraus2020scared}. Lee and See (\citeyear{lee2004trust}) defined trust in automation as the "attitude that an agent will help achieve an individual's goals in a situation characterized by uncertainty and vulnerability." Traditionally, trust research has emphasized cognitive factors, such as system reliability and performance assessments. However, recent work underscores the significant, and potentially dominant, role of affective components in shaping trust-related behaviors \citep{madsen2000measuring, de2020towards}. Within time-constrained and high-risk environments such as driving, emotions can serve as rapid, intuitive signals that influence whether individuals choose to rely on or avoid automation. Despite this recognized importance, the specific ways in which emotions mediate the relationship between risk perception, system performance, and trust in AVs remain underexplored.

Understanding the interplay between emotion, risk, and trust is crucial for achieving appropriate trust calibration – the alignment of a user's subjective trust with the objective capabilities of the AV \citep{de2020towards}.  Miscalibration, whether overtrust or undertrust, can lead to suboptimal outcomes, including accidents or inefficient system use. Prior research began to categorize emotional responses in human-automation interaction, revealing that negative emotions like anxiety and discomfort tended to have a disproportionately strong negative impact on trust, while positive emotions like confidence had a weaker, slower-building positive effect \citep{buck2018user, lewandowsky2000dynamics}.  However, these findings, often derived from contexts like computer warnings, may not fully generalize to the dynamic, high-risk environment of  automated driving.

Building upon the three-layered trust model (dispositional, learned, and situational trust) proposed by Hoff and Bashir (\citeyear{hoff2015trust}), this study addresses a critical gap in the literature: the mediating role of specific emotions in the relationship between risk perception, AV system performance, and driver trust. While much research focused on cognitive aspects of trust and performance, less is known about how affective responses dynamically influence trust calibration during real-time AV interactions, particularly when faced with unexpected system behaviors.

To address this gap, the current study investigates the impact of risk perception, AV performance (with and without errors), and emotional responses on trust in automated vehicles. Specifically, this research aims to answer the following research questions:

\begin{itemize}
    \item RQ1: How does risk perception affect drivers' trust in AVs under varying performance conditions (i.e., with error vs. without error)?
\end{itemize}
\begin{itemize}
    \item RQ2: What is the impact of affective responses, such as anxiety or confidence, on the calibration of trust in AVs during automated driving, and how do these emotions mediate the relationship between perceived risks and overall trust in AVs?
\end{itemize}

By addressing these questions, the study seeks to provide a more nuanced understanding of how trust is formed, maintained, and potentially eroded in the context of automated driving, ultimately contributing to the design of safer and more trustworthy AV systems.

\section{Related Work}
Trust in automation has been extensively studied, with much of the early research emphasizing cognitive evaluations of system performance, such as reliability, predictability, and transparency \citep{lee2004trust, hancock2011meta}. This cognitive trust is built through rational assessments of an automated system's capabilities and the user's expectations regarding its consistent behavior. However, as automation becomes more sophisticated and integrated into safety-critical contexts like automated driving, the role of affective responses, i.e. emotions experienced in interaction with the system, is increasingly recognized as a critical, and potentially dominant, factor in shaping trust \citep{choi2015investigating, de2020towards}. A comprehensive understanding of trust in AVs requires integrating both cognitive and affective dimensions.

\subsection{Cognitive and Affective Dimensions of Trust in Automation}
\subsubsection{Cognitive Trust: The Rational Assessment of Performance}
Cognitive trust is often conceptualized as an evaluative, logical process. Users rationally assess a system's functionality, performance, and adherence to expected outcomes \citep{lee1992trust}. Hoff and Bashir \citeyearpar{hoff2015trust}'s three-layered model of trust provides a useful framework for understanding this cognitive dimension:

\begin{itemize}
    \item Dispositional Trust:  An individual's general tendency to trust automation, shaped by personality and prior experiences.
\end{itemize}
\begin{itemize}
    \item Learned Trust:  Trust formed through direct experience with a specific system, adjusted based on observed performance.
\end{itemize}
\begin{itemize}
    \item Situational Trust:  The dynamic, context-dependent aspect of trust, influenced by factors like environmental conditions, task criticality, and perceived risk.
\end{itemize}

Within this framework, cognitive trust is built incrementally as users gather information about a system's capabilities and reliability.  Assessments of predictability, reliability, and accuracy are essential to establish this form of trust \citep{dzindolet2003role, ayoub2021investigation}. Users are more likely to trust an automated system when they understand how it works and when it consistently meets performance expectations.

\subsubsection{Affective Trust: The Role of Emotional Responses}
While cognitive evaluations are essential, research increasingly highlights the influence of emotional responses on trust, often independent of, or even overriding, rational assessments \citep{madsen2000measuring}. Madsen and Gregor \citeyearpar{madsen2000measuring} were among the first to emphasize the role of emotions like comfort, anxiety, and fear in shaping trust in automation.  These emotional experiences and implicit emotional cues can significantly impact a user's willingness to rely on a system.
For instance, a user might experience anxiety or discomfort when interacting with a new or highly automated system like an AV, leading to lower trust regardless of the system's objective performance. In contrast, positive emotional experiences, such as smooth and uneventful interactions, can foster greater trust. Choi et al. \citeyearpar{choi2015investigating} emphasized that affective trust could operate independently of actual reliability; users might continue to trust or distrust a system based on their emotional responses, even when cognitive assessments suggested otherwise. Lee and See \citeyearpar{lee2004trust} explicitly argued the affective process could not be neglected in trust and that trust might not be such a profound factor without emotions.

\subsubsection{Risk Perception and its Influence on Trust}
The perception of risk is a key factor influencing trust in automated systems, particularly in complex, high-risk situations like  automated driving. Perceived risk, the subjective evaluation of potential harm or failure, can lead to mistrust or even distrust in automation, even when the system demonstrates objective reliability \citep{slovic1987perception, parasuraman1997humans}.

Several factors influence perceived risk in automation, including system transparency, controllability, and predictability \citep{hoff2015trust}.  A highly autonomous system with limited user control may be perceived as riskier, especially when users cannot intervene during malfunctions or unexpected behaviors \citep{muir1994trust}.  It is important to distinguish between actual risk (objective probability of failure) and perceived risk, which is shaped by personal biases, prior experiences, and external influences, such as media portrayals \citep{shariff2017psychological}.

In the context of AVs, risk perception is a key determinant of public trust and adoption \citep{choi2015investigating}. Individuals may perceive AVs as inherently riskier due to complexity or unfamiliarity, even when empirical evidence suggests that automation can enhance safety.  Users often focus on worst-case scenarios, such as fatal accidents caused by system errors, even if these events are statistically rare \citep{gkartzonikas2022tale}.  Risk communication also plays a significant role.  Information about AV risks, particularly scenarios highlighting potential failures, can lower trust and increase reluctance to use AVs, even when presented alongside data demonstrating overall safety benefits \citep{bansal2016assessing, cunningham2019public}.

\subsection{Emotion as a Mediator of Risk Perception and Trust}
Emotions are essential for understanding the relationship between risk perception and trust. This means that while cognitive assessments of risk and system performance are important, the emotional reactions these assessments trigger significantly shape the development and calibration of trust. This mediation is particularly relevant in AV contexts, where high uncertainty and perceived risk can evoke strong emotional responses \citep{hohenberger2016and}.

When users perceive a high risk of malfunction or failure in an AV, they may experience fear or discomfort, which directly reduces trust, even if a cognitive evaluation might suggest the vehicle is reliable \citep{choi2015investigating}. In contrast, positive experiences, such as flawless performance or effective handling of complex driving situations, can reduce anxiety and foster confidence, mediating the formation of higher levels of trust \citep{merat2009drivers}.  Negative performance or even anticipation of risk can lead to emotions such as stress or anxiety, which mediate lower levels of trust \citep{de2020towards}. The affect heuristic \citep{slovic1987perception} supports this idea, proposing that emotions often guide risk and benefit judgments, aligning with the findings that emotional reactions are key drivers of trust in AVs \citep{shariff2017psychological}.

The emotions of users have a pivotal impact on how they see and trust automated systems, especially in the AV where they can have accidents. Emotional responses can affect the willingness of users to trust, such as anxiety, fear, and confidence \citep{choi2015investigating}. When the AV works without problems, it can reduce anxiety and fear, which as a result would increase the confidence and trust of users in the system \citep{gold2013take}. The unpredictability of the AV can cause the user to be more anxious, as users are sensitive to system errors \citep{merritt2013trust}. Kraus et al. \citeyearpar{kraus2020scared} found that higher state anxiety at the introduction of an automated driving system was associated with lower trust in it. This anxiety was also predicted by lower self-esteem and self-efficacy, as well as higher levels of depressiveness, suggesting that users with negative self-evaluations tended to be more anxious and consequently less trusting of new automated driving technology.

\subsection{Current Study}
While significant progress has been made in understanding cognitive trust and risk perception in automation, a crucial gap remains: a systematic investigation of how specific emotions mediate the relationship between risk perception, AV performance, and trust calibration.  Much of the existing literature focuses on the influence of rational assessments of reliability and transparency \citep{lee2004trust, hoff2015trust}, but fewer studies have explicitly explored the affective dimension, particularly the mediating role of emotions triggered by perceived risks and system errors.  Furthermore, few studies have used a factor-based approach to systematically dissect the contributions of different emotional components to trust formation and calibration in the dynamic context of AV driving. This study aims to address this gap by integrating cognitive, affective, and (potentially) psychophysiological data to provide a more comprehensive understanding of trust in AVs.

\section{Methods}
\subsection{Participants}
Seventy-five participants were recruited via university email and compensated with \$20 USD for their participation.  The sample comprised 46 male and 29 female, with ages ranging from 18 to 60 years (M = 25.1, SD = 5.83). All participants held a valid driver's license and reported normal or corrected-to-normal vision (without the need for glasses, as contact lenses are generally acceptable for driving and eye-tracking).  The study protocol was approved by the Institutional Review Board of the University of Michigan (HUM00215101). Data from five participants were excluded from the final analysis due to technical issues. One participant was excluded due to hyperhidrosis, which interfered with the reliable measurement of physiological signals.  Four additional participants were excluded due to sensor calibration errors.  This resulted in a final sample of 70 participants.

\subsection{Apparatus}
The experiment was carried out with a desktop computer setup when participants watched real-life driving scenarios presented on a 27-inch LCD display with a resolution of 1920 x 1200 pixels. 
Physiological responses were recorded using a Shimmer3 GSR+ Unit (Shimmer, MA, USA). This device measured skin conductance responses and heart rate at a sampling rate of 128 Hz. 
The iMotions software platform (imotions.com, MA, USA) was used to synchronize the eye-tracking and physiological data streams, allowing for precise temporal alignment of these measures with the events in the driving scenarios.  Self-reported measures (described below) were collected using the Qualtrics online survey platform (qualtrics.com, Seattle, WA).

\subsection{Experiment Design}
This study employed a mixed-subjects design, with risk perception as a between-subjects factor and AV performance as a within-subjects factor. Participants were randomly assigned to one of three risk perception conditions: \emph{high-risk}, \emph{low-risk}, or a \emph{control} condition. Within each risk perception condition, participants experienced two AV performance scenarios: one \emph{with system errors} and one \emph{without system errors} (see Table \ref{tab:PerformanceManipulation}). To account for potential biases due to presentation order, the performance conditions (with error vs. without error) were counterbalanced, where each participant experienced the conditions in a randomly generated order.

\subsubsection{Independent Variables:}

\begin{itemize}
    \item[] \textbf{Risk Perception (Between-Subjects):} We manipulated risk perception through informational videos presented before the driving scenarios (see Table \ref{tab:RiskManipulation} and Procedure for details).
    \begin{itemize}
        \item \emph{High-Risk:} Emphasized potential AV vulnerabilities, operational limitations, and accident scenarios.
        \item \emph{Low-Risk:} Highlighted the safety benefits and accident-prevention capabilities of AVs.
        \item \emph{Control:} No risk-related information was provided.
    \end{itemize}

    \item[] \textbf{AV Performance (Within-Subjects):}
    \begin{itemize}
        \item \emph{With Errors:} The AV exhibited specific errors, including inability to navigate a construction zone, incorrect guidance from a remote support team, and a lack of clear explanation for disengagement (see Table \ref{tab:PerformanceManipulation}).
        \item \emph{Without Errors:} The AV demonstrated flawless performance, adhering to traffic laws and exhibiting smooth, confident navigation (see Table \ref{tab:PerformanceManipulation}).
    \end{itemize}
\end{itemize}

\subsubsection{Dependent Variables:}
This study used both subjective and objective measures. The subjective dependent variables were situational trust, perceived risk, and emotions. From objective measures, physiological arousal, including Galvanic Skin Response (GSR) and heart rate (HR), was used in this study.

\emph{Trust} in the AV was measured using the Situational Trust Scale for Automated Driving (STS-AD) \citep{holthausen2020}, with a six-item scale (i.e., trust, performance, non-driving-related tasks, risk, judgment, and reaction).   
 
\emph{Perceived Risk} was measured based on their perceived driving experience in four dimensions (reliability, accident risk, performance uncertainty, and safety) \citep{zhang2019roles}. 
Participants rated their trust and perceived risk on a seven-point Likert scale (1 = Strongly Disagree, 7 = Strongly Agree) after each driving scenario.

\emph{Emotional Responses:} To assess participants' emotional states, we employed a 19-item emotion word list (i.e., confident, secure, grateful, happy, respectful, disdainful, scornful, contemptuous, hostile, resentful, ashamed, humiliated,  nervous, anxious, confused, afraid, freaked out, lonely, isolated) \citep{jensen2020anticipated}. This list draws upon discrete emotion theory and is grounded in affective neuroscience, reflecting both analytical and syncretic-affective dimensions of emotional experience \citep{chaudhuri2006emotion}. The selection of these specific emotion words was informed by their prior use in research examining the role of emotions in decision-making processes and the development of trust in human-AV interactions \citep{avetisian2022anticipated}. Participants provided subjective ratings for each emotion word using a seven-point Likert scale, indicating the extent to which they experienced each emotion during each drive.

\emph{Physiological Arousal:} GSR and HR measures were continuously monitored throughout the study as objective indicators of physiological arousal and emotional intensity. Our data processing procedure involved multiple steps to ensure high-quality analysis. First, we cleaned the raw data by removing noise and addressing missing values using forward linear interpolation to estimate missing and invalid data points and ensure temporal continuity in the physiological signals. To reduce random fluctuations while preserving meaningful physiological responses to experimental stimuli, we further calculated moving averages for both GSR and HR using 10-second window intervals. For the GSR data, we extracted phasic (skin conductance response (SCR)) and tonic (skin conductance level (SCL)) components using the Neurokit2 tool \citep{Makowski2021neurokit}. The phasic component was used to identify rapid, event-related physiological reactions to stimuli, while the tonic component reflected the overall arousal state and baseline of the participant.

\subsection{Procedure}

The experiment followed a structured protocol designed to examine the relationships between risk perception, AV performance, emotional responses, and trust. 

\emph{Briefing and Consent:} Participants received a briefing explaining the study's goals, procedures, and potential risks. They provided written informed consent after understanding the nature of the experiment, including physiological data collection.
\emph{Device Setup and Baseline:} For GSR data collection, snap electrodes were attached to the participant's non-dominant (left) palm, and a photoplethysmography (PPG) sensor was placed on the left earlobe to measure HR. To establish a baseline representing participants' typical resting state, with minimal arousal, a one-minute physiological recording was taken. This recording was used to account for individual physiological differences.

\emph{Demographics and Dispositional Trust:} Participants completed a demographic (age, gender, driving experience, technology familiarity) and general trust questionnaires to assess dispositional trust in AVs. This controlled for individual differences in pre-existing trust tendencies \citep{Ayoub2021Modeling}.

\emph{Risk Perception Manipulation:} Participants were randomly assigned to one of three risk perception conditions (high-risk, low-risk, or control). Participants in high and low-risk groups received an introduction to AVs designed to either enhance or reduce perceived risk, followed by a corresponding informational video. (see Table \ref{tab:RiskManipulation}).


\begin{table}[H]
\centering
\small
\caption{Information script to videos used for risk manipulation.}
\renewcommand{\arraystretch}{0.8}
\begin{tabularx}{\textwidth}{p{0.15\linewidth}p{0.8\linewidth}}
\hline
Risk & Introduction script and video links\\ 
\hline
High & Now, self-driving vehicles have a higher rate of accidents compared to human-driven vehicles, but the injuries are less serious. On average, there are 9.1 self-driving vehicle accidents per million miles driven, while the same rate is 4.1 crashes per million miles for human-driven vehicles. Self-driving vehicles had a higher rate of injury per crash: 0.36 injuries per crash, compared with 0.25 for human-driven vehicles.
\href{https://youtu.be/RC9iK1lV77E?si=8oByKl_I9IPdV6E2}{\textcolor{blue}{video link}}\\
Low &The safety benefits of self-driving vehicles are paramount. Self-driving vehicles’ potential to save lives and reduce injuries is rooted in one critical and tragic fact: 94\% of serious crashes are due to human error. Self-driving vehicles have the potential to remove human error from the crash equation, which will help protect drivers and passengers, as well as bicyclists and pedestrians. When you consider more than 35,000 people die in motor vehicle-related crashes in the United States each year, you begin to grasp the lifesaving benefits of driver assistance technologies. \href{https://youtu.be/O4IUc0xXZqo?si=xiuvbPQZ8-x8cXmx}{\textcolor{blue}{video link}}\\
\hline
\end{tabularx}
\label{tab:RiskManipulation}
\end{table}

After the video (or no exposure in the control condition), participants rated their \emph{initial} learned trust in AVs using a ten-item questionnaire \citep{avetisyan2024building}.

\emph{AV Performance Scenarios:} Participants viewed two pre-recorded videos of real-world cases of AV driving scenarios, presented in a counterbalanced order. In the \emph{``with errors''} condition, the AV exhibited failures in navigating a construction zone, responding to remote support, and providing clear explanations (\href{https://youtu.be/z3kTQ3HB4xc?si=DtZvPC7LoEvxVBIs}{video link}).
In the \emph{``without errors''} condition, the AV demonstrated flawless performance, navigating smoothly and confidently (\href{https://youtu.be/P_uixyQLeFo?si=HyS4ZpqgJeRlNaH7}{video link}).

Physiological responses (GSR and HR) were continuously recorded throughout both scenarios. Immediately after each driving scenario, participants rated their trust, perceived risk, and emotional responses, with a one-minute break between drives.

The entire experiment was conducted in a controlled laboratory setting, maintaining consistent lighting, sound, and temperature to minimize extraneous influences on physiological responses \citep{balters2017capturing}.

\begin{table}[H]
\centering
\small
\caption{Information script and links to videos used for performance manipulation.}
\renewcommand{\arraystretch}{0.8}
\begin{tabularx}{\textwidth}{p{0.15\linewidth}p{0.8\linewidth}}
\hline
AV Performance & Description and video links\\ 
\hline
With Error & The car's inability to navigate a construction zone points to software or mapping data limitations, indicating a higher risk in dynamic environments with unexpected obstacles or changes. The car required intervention due to incorrect guidance provided by the remote fleet response team, which indicates a potential risk of human error within the support system, leading to dangerous situations for passengers. The car doesn't clearly explain the reason for the disengagement and the need for support. This lack of transparency raises concerns about the system's overall reliability and ability to handle unforeseen circumstances, increasing the potential risk for passengers
(\href{https://youtu.be/z3kTQ3HB4xc?si=DtZvPC7LoEvxVBIs}{\textcolor{blue}{video link}}).\\
Without Error &The car demonstrated confident navigation, adhering to traffic laws, and exhibiting smooth acceleration, braking, and turning while demonstrating awareness of parked cars, pedestrians, and other vehicles. Both the in-car display and the Waymo app provided clear information and support options to the passenger. The ride was completed faster than anticipated, offering a quiet and spacious environment for the passenger
(\href{https://youtu.be/P_uixyQLeFo?si=HyS4ZpqgJeRlNaH7}{\textcolor{blue}{ video link}}).\\\hline
\end{tabularx}
\label{tab:PerformanceManipulation}
\end{table}

\section{Data Analysis}
The data analysis proceeded in three main stages: (1) preliminary analyses to validate manipulations and assess main effects, (2) exploratory factor analysis (EFA) to reduce the dimensionality of the emotional response data, and (3) mediation analysis to test the hypothesized relationships between risk perception, AV performance, emotional responses, and trust.

\subsection{Preliminary Analyses: Manipulation Checks and Main Effects}
To validate the effectiveness of the risk perception manipulation, we conducted a one-way analysis of variance (ANOVA) with significance level at 0.05. The analysis compared learned trust measurements, collected immediately after the risk manipulation videos, across the three risk perception conditions (high-risk, low-risk, and control). We hypothesized that participants exposed to the high-risk condition would report significantly lower trust than those in either the low-risk or control conditions.

To examine the main effects of risk perception and AV performance on trust and physiological measures (HR and SCR), a series of mixed-design ANOVAs were conducted.  For trust, the dependent variable was the post-scenario trust ratings (STS-AD). For the physiological measures, the dependent variables were the mean HR  and mean SCR during each scenario.
Significant main effects and interactions were followed up with post-hoc Tukey's HSD tests to determine the nature of the differences.

\subsection{EFA of Emotional Responses}

To reduce the dimensionality of the 19 self-reported emotion items and identify underlying latent emotional constructs, an EFA was performed, as we anticipated that the underlying emotional factors might be correlated. The number of factors to retain was determined based on multiple criteria, including Kaiser's criterion (eigenvalues > 1), scree plot examination, and parallel analysis. Items with factor loadings of 0.40 or greater on a given factor were considered to meaningfully contribute to that factor. The resulting factors were labeled based on the substantive meaning of the items loading onto them.

\begin{figure}
    \centering
    \includegraphics[width=1\linewidth]{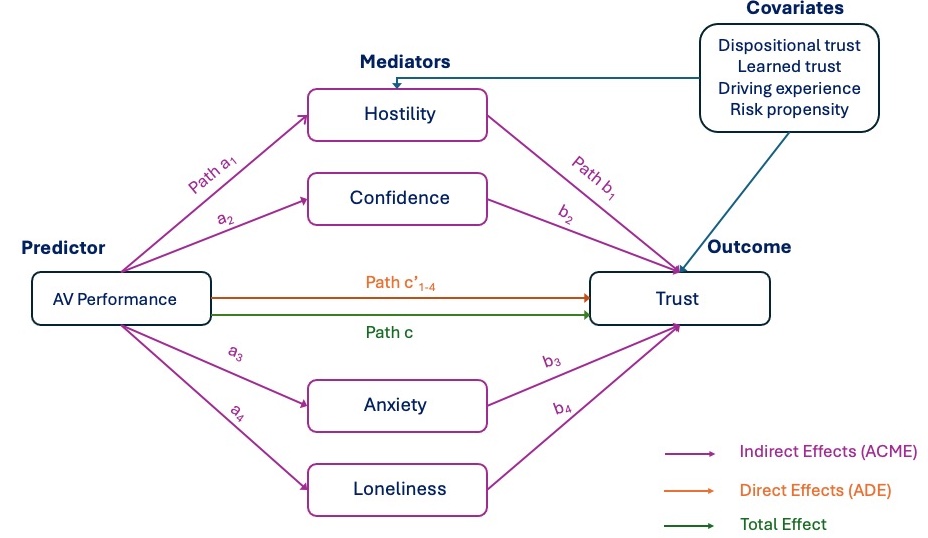}
    \caption{Mediating relationship between AV Performance,  emotions, and trust. ACME stands for Average Causal Mediation Effects, ADE stands for Average Direct Effects, Total Effect is a sum of a mediation (indirect) effect and a direct effect.}
    \label{fig:mediation}
\end{figure}

\subsection{Mediation Analysis}

To test the hypothesized mediating role of emotional responses in the relationship between risk perception, AV performance, and trust, we employed a mediation analysis framework following the recommendations of Baron and Kenny \citeyearpar{baron1986moderator}. Given our repeated-measures design, this approach involved a series of linear mixed-effects (LME) models, implemented in R  using the \texttt{lme4} \citep{bates2015package} and \texttt{mediation} \citep{tingley2014mediation} packages. 
For each emotional factor identified through the EFA, we conducted the following mediation analysis:
\begin{enumerate}
    \item \textbf{Path a (IV → Mediator):}  An LME model was constructed with the emotional factor as the dependent variable, risk perception and AV performance as fixed effects, and participant ID as a random effect.  This model assessed whether risk perception and AV performance significantly predicted the emotional factor.  We also included an interaction between Risk Perception and AV Performance.
    \item \textbf{Path b (Mediator → DV, controlling for IVs):} An LME model was constructed with trust (STS-AD score) as the dependent variable, the emotional factor as a fixed effect, risk perception, and AV performance as fixed effects, and participant ID as a random effect. This model assessed whether the emotional factor significantly predicted trust, controlling for the direct effects of risk perception and AV performance.
    \item \textbf{Path c (Direct Effect: IV → DV):} An LME model was constructed with trust (STS-AD score) as the dependent variable, risk perception and AV performance as fixed effects, and participant ID as a random effect. This model assessed the direct effect of the independent variables on trust, without the mediator.
    \item \textbf{Path c' (Indirect Effect):} The indirect effect (a * b) represents the extent to which the emotional factor mediates the relationship between the independent variables and trust. 
\end{enumerate}

To control for potential confounding variables, we included the following covariates in all LME models: dispositional trust, learned trust, driving experience, and risk propensity. These covariates were selected based on prior research suggesting their potential influence on trust in automation. All statistical analyses were conducted using R version 4.4.2.  Statistical significance was set at $\alpha = 0.05$ for all analyses.

\section{Results}
\subsection{Manipulation check}
\begin{figure}
    \centering 
    \begin{subfigure}[b]{0.43\textwidth} 
        \centering 
        \includegraphics[width=\textwidth]{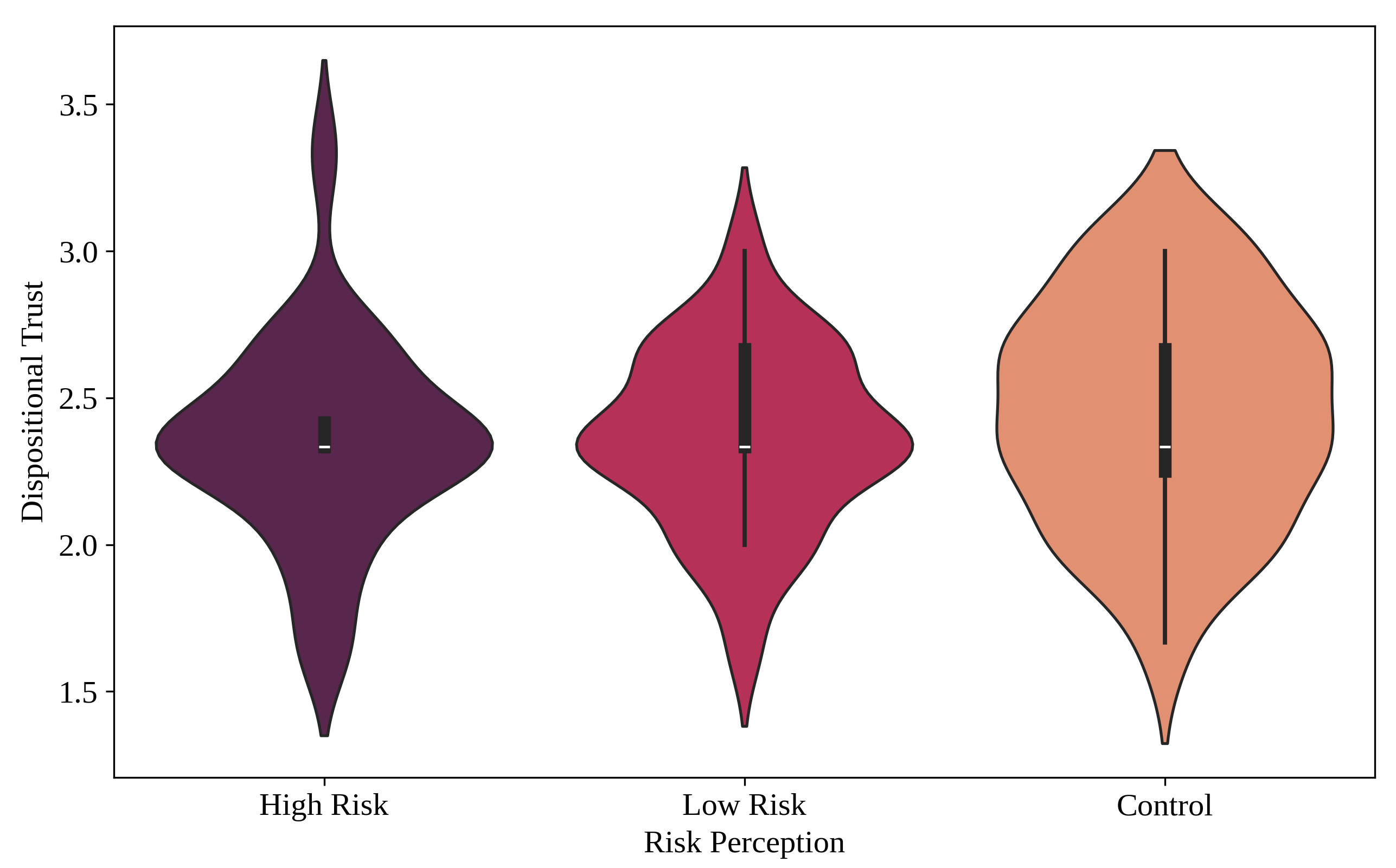} 
        \caption{Dispositional Trust} 
        \label{fig:dispTrust} 
    \end{subfigure} 
    \hspace{1cm}
    \begin{subfigure}[b]{0.43\textwidth} 
        \centering 
        \includegraphics[width=\textwidth]{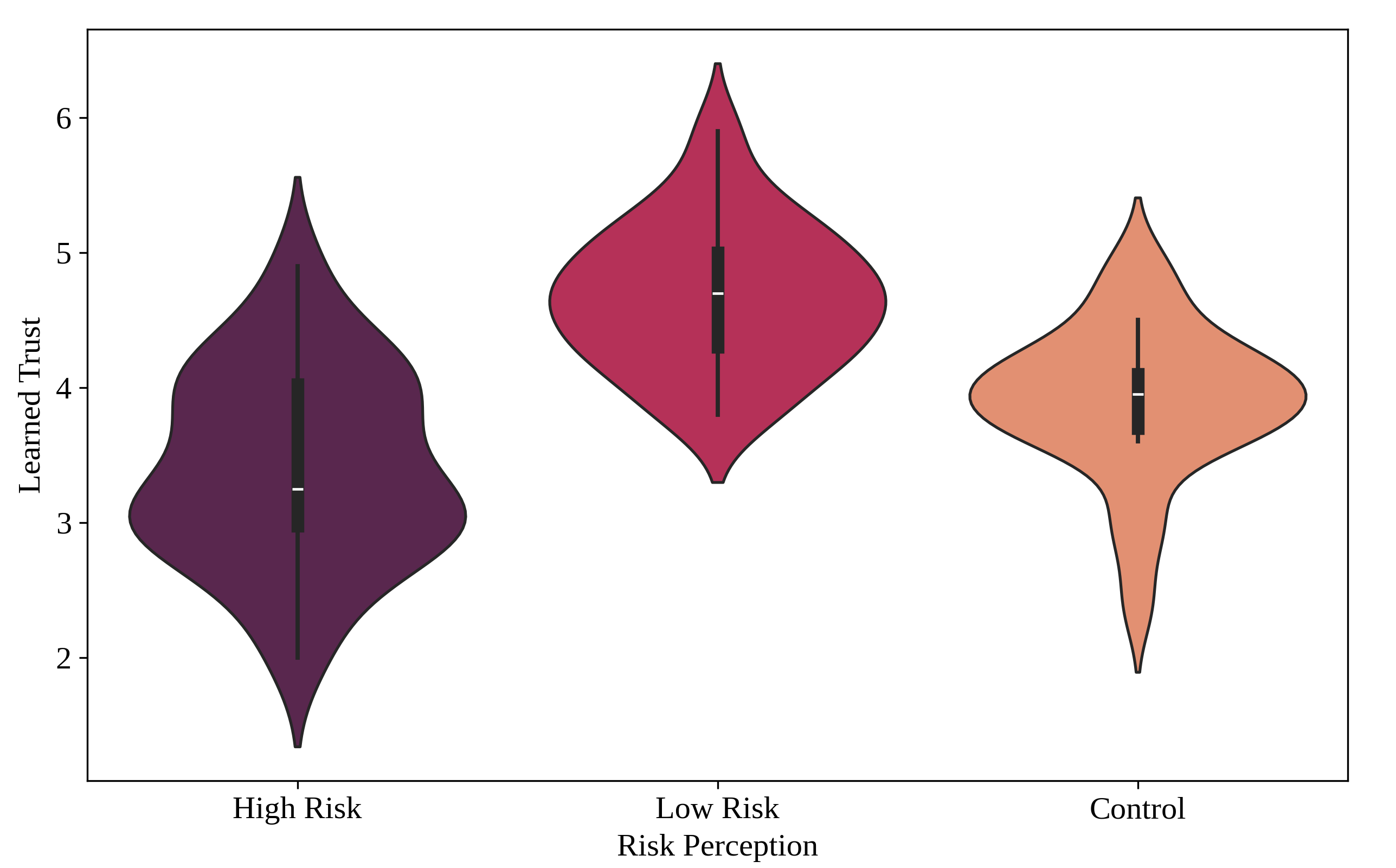}
        
        \caption{Learned Trust} 
        \label{fig:learnedTrust} 
    \end{subfigure} 
    \caption{Comparison of trust levels in AV before (Dispositional Trust) and after (Learned Trust) exposure to risk-related information.}
 
    \label{fig:initaltrust}
\end{figure}

In this study, the effect of the provided information was explored by measuring the learned trust ratings after manipulation. The analysis of dispositional trust confirmed no initial differences among conditions (high, low, control) ($F(2, 67) = 0.54$, $p < .585$, $\eta^2 = 0.016$). In contrast, a one-way ANOVA result showed a significant main effect of risk perception ($F(2, 67) = 24.94$, $p < .001$, $\eta^2 = 0.426$), demonstrating that our manipulation accounted for approximately 43\% of the variance in learned trust scores. 
Post-hoc test indicated that participants in the high-risk condition reported significantly lower learned trust than those in both the low-risk ($p < .001$) and control ($p = .017$) conditions.  Furthermore, the low-risk condition resulted in significantly higher learned trust than the control condition ($p < .001$). These results confirmed that the risk manipulation successfully influenced participants' learned trust in the intended directions. Figure \ref{fig:initaltrust} shows the comparison of trust levels measured before the manipulation (dispositional trust; Figure \ref{fig:dispTrust}) and after exposure to the risk perception conditions (learned trust; Figure \ref{fig:learnedTrust}), illustrating the differential impact of our experimental manipulation on trust formation.

A two-way repeated-measures ANOVA was then conducted on post-scenario perceived risk ratings. Contrary to expectations, there was no significant main effect of risk perception condition ($F(2, 67) = 0.59$, $p = .555$, $\eta^2 = 0.017$).  However, there was a significant main effect of AV performance ($F(1, 67) = 81.28$, $p < .001$, $\eta^2 = 0.548$), indicating that perceived risk was significantly higher in the "with errors" condition compared to the "without errors" condition.  The interaction between risk perception and AV performance was not significant ($F(2, 67) = 0.28, p = 0.759, \eta^2 = 0.008$). 

Although ANOVA did not reveal a significant interaction, descriptive trends (see Table \ref{tab:descript}) suggested a potential moderating role of risk information. In both AV performance conditions, the low-risk group tended to report lower perceived risk than the control and high-risk groups. The high-risk group generally reported slightly higher perceived risk than the control group. Notably, the control group's perceived risk increased more dramatically in the "with errors" condition compared to the high-risk group, suggesting that pre-existing risk information might buffer the impact of errors on risk perception.

\begin{table}[H]
   \centering
   \caption{Mean values (SD) for dependent measures by AV performance and risk perception conditions}
   \label{tab:descript}
   \begin{adjustbox}{width=.9\textwidth}
   \begin{tabular}{lccccccc}
        \hline
      Performance& Risk Perception& Risk & Trust & Hostility & Confidence & Anxiety & Loneliness \\
       \hline
       \multirow{3}{*}{With Errors} 
       & Control & 5.40(0.83) & 3.39(0/95) & 3.42(1.33) & 2.97(1.25) & 4.61(1.45) & 3.06(1.87) \\
       & High Risk & 5.22(1.18) & 3.22(0.70) & 3.65(1.29) & 3.08(1.42) & 4.39(1.26) & 2.65(1.51) \\
       & Low Risk & 4.99(1.07) & 3.49(0.96) & 2.72(1.28) & 3.25(1.38) & 4.24(1.37) & 2.93(1.83) \\
       
       \multirow{3}{*}{Without Errors}
       & Control & 3.86(1.42) & 5.18(0.89)  & 1.68(0.91) & 5.16(1.12) & 3.05(1.19) & 2.50(1.55) \\
       & High Risk & 3.94(1.29) & 5.12(0.61) & 1.80(0.85) & 4.95(0.99) & 3.02(1.23) & 1.85(0.99) \\
       & Low Risk & 3.66(1.36) & 5.35(0.55) & 1.42(0.49) & 5.31(0.99) & 3.07(1.27) & 2.66(1.64) \\
       \hline
   \end{tabular}
   \end{adjustbox}
\end{table}

\subsection{Exploratory Factor Analysis}
An EFA was conducted on the 19 emotion items to identify the underlying emotional constructs.  Kaiser-Meyer-Olkin (KMO) measure of sampling adequacy was 0.881, indicating that the sample was suitable for factor analysis. The Bartlett test for sphericity was significant ($\chi^2(171) = 2242.94, p < .001$), confirming that the correlation matrix was appropriate for factor extraction. The EFA revealed four distinct emotional factors: Hostility (i.e.,Disdainful, Scornful, Resentful, Contemptuous, Ashamed, Humiliated), Confidence (i.e., Happy, Grateful, Confident, Secure, Respectful), Anxiety (i.e.,Anxious, Afraid, Nervous, Freaked out), and Loneliness (i.e., Isolated, Lonely), which are consistent with previous findings \citep{buck2018user,avetisian2022anticipated}. These factors explained a cumulative 0.7363 of the total variance, with hostility accounting for the largest portion (0.454). As shown in Table \ref{tab:Factor Analysis}, the internal consistency of the factors, measured by Cronbach's $\alpha$, was strong for hostility ($\alpha$ = 0.914), confidence ($\alpha$ = 0.941) and anxiety ($\alpha$ = 0.833), while loneliness exhibited a relatively lower consistency ($\alpha$ = 0.599).

\begin{table}[H]
\centering
\small
\caption{Factor loadings for each of the 19 emotion items. Only loadings above 0.4 are shown. Scale reliability with Cronbach’s $\alpha$, and the average inter-item correlations for each factor are shown at the bottom of the table.}

\renewcommand{\arraystretch}{0.8}
\begin{tabularx}{\textwidth}{p{0.18\linewidth}p{0.15\linewidth}p{0.15\linewidth}p{0.15\linewidth}p{0.15\linewidth}}
\hline
 &Hostility&Confidence& Anxiety&Loneliness\\ 
\hline 
Disdainful&  0.676&  &  & \\
Scornful &  0.814&  &  & \\
Contemptuous  &  0.751&  &  & \\
Hostile &  0.728&  &  & \\
Resentful &  0.754&  &  & \\
Ashamed&  0.695&  &  & \\
Humiliated    &  0.648&  &  & \\ 
Confident&  &  0.765&  & \\
Secure&  &  0.751&  & \\
Grateful&  &  0.881&  & \\
Happy        & & 0.883& &\\
Respectful  & & 0.759& &\\
Nervous      & & & 0.749&\\
Anxious     & & &  0.813&\\
Confused    & & & 0.444&\\
Afraid       & & &  0.764&\\
Freak-out   & & & 0.624&\\
Lonely        & & &  &0.785\\
Isolated     & & & &0.858\\
$\alpha$ & 0.914& 0.941&  0.599& 0.833\\
IIC& 0.941& 0.765& 0.216&0.072\\
 
\hline
\end{tabularx}
\label{tab:Factor Analysis}
\end{table}

\subsection{The effects of Risk Perception and AV Performance on Trust}
To investigate the relationships between risk perception and AV performance on trust, we employed a LME regression model while controlling for individual differences such as driving experience, risk propensity, and both dispositional and learned trust. The model of risk perception revealed that it had no significant effect on trust ($\beta=-0.04, p=0.745$) or emotion factors and was consequently excluded from further analysis. In contrast, AV performance exhibited a significant positive effect on trust ($\beta=1.84, p<0.001$), indicating that performance without errors significantly enhanced trust levels. 
Regarding the covariates, both dispositional and learned trust showed positive associations with trust. Learned trust demonstrated a significant positive effect ($\beta=-0.249, p<0.01$), suggesting that participants' previous experiences with automated systems positively influenced their trust levels. Dispositional trust ($\beta=0.382, p=0.052$) and experience in driving ($\beta=-0.04, p=0.065$) showed marginally significant effects, indicating that individuals' general tendency to trust others may positively contribute to their trust in AVs. 
Risk propensity also showed a marginally significant positive relationship with trust ($\beta=0.109, p=0.054$), suggesting that individuals with higher risk tolerance might be more inclined to trust AVs. Years of driving experience showed a slight negative association ($\beta=-0.043, p=0.065$), indicating that more experienced drivers might be somewhat more cautious in their trust of AVs.

\subsection{Mediation Analysis: The Role of Emotions in Trust}
To examine the potential mediating effects, we constructed mediation models (illustrated in Figure \ref{fig:mediation}) based on the significant relationships identified in our previous analyses. For each pathway in the mediation model, we conducted a separate LME regression analysis, with the  results presented in Table \ref{tab:mixed-regression}.

We constructed four separate LME regression models examining the relationship between AV performance and each emotional factor (Path $a_1-a_4$) to investigate the emotional mediation of AV performance on trust development. Subsequent analyses revealed that AV performance had significant effects across all emotional dimensions. Specifically, error-free performance significantly reduced hostility ($\beta=-1.64, p<0.001$) and alleviated anxiety-related emotions ($\beta=-1.38, p<0.001$) toward the system. Furthermore, error-free performance notably decreased loneliness ($\beta=-0.55, p=0.039$), suggesting that participants felt less isolated when the AV operated smoothly.
Finally, error-free AV performance substantially increased confidence ($\beta=2.03, p<0.001$), signifying that operation without errors increased participants' confidence in the AV system.
We deployed additional models to quantify the total effect of AV performance on trust and examine the mediating effect of each emotional factor on the AV performance-trust relationship.
The results confirm that AV performance continued to have a significant direct effect on trust ($p<0.001$ for all models), even when emotions were included as mediators. 
The analysis of individual emotional mediators revealed distinct patterns in how AV performance influences trust (see Table \ref{tab:mixed-regression}). Confidence emerged as a strong positive predictor of trust ($\beta = 0.87, p < 0.001$), mediating 46.7\% of the total effect of AV performance on trust (ACME: $\beta = 0.87, p < 0.001$). Although the direct effect remained significant after accounting for confidence (ADE: $\beta = 0.98, p < 0.001$), indicating partial mediation, these findings suggest that improved AV performance substantially increased confidence, which led to increased trust in the system.
Hostility demonstrated a significant negative relationship with trust ($\beta = -0.13, p = 0.028$), mediating 11.3\% of the total effect (ACME: $\beta = 0.21, p = 0.036$). The majority of the influence occurred through the direct pathway (ADE: $\beta = 1.64, p < 0.001$), suggesting that while AV performance partially reduces hostility-related emotions, its primary impact on trust operates through other mechanisms.
Similarly, anxiety exhibited a significant negative effect on trust ($\beta = -0.24, p < 0.001$), with well-functioning AVs appearing to reduce anxiety-related emotions, thereby promoting greater trust. This factor partially mediated 17.7\% of the AV performance effect (ACME: $\beta = 0.31, p < 0.001$), with the direct effect remaining substantial (ADE: $\beta = 1.53, p < 0.001$).
Loneliness, while a significant predictor of trust ($\beta=0.09$, $p=0.008$), did not mediate the relationship between AV performance and trust (ACME: $\beta = 0.01, p = 0.498$). These results indicated that while loneliness may independently influence trust formation in AVs, it does not function as a mediating pathway in this relationship.

\begin{table}[H] 
    \centering
    \caption{LME Regression results grouped by Mediator variables.}
    \begin{tabular}{lccccc}
        \toprule
        \textbf{Path} & \textbf{Coef ($\beta$)} & \textbf{SE} & \textbf{$p$} & \textbf{CI [2.5\%]} & \textbf{CI [97.5\%]}\\
        \midrule
        \textbf{Total Effect} & 1.85 & 0.12 & 0.000 & 1.61 & 2.09 \\
        \midrule
        \textbf{Hostility} & & & & & \\
        AV Performance → Hostility & -1.64 & 0.18 & 0.000 & -1.97 & -1.28 \\
        Hostility → Trust & -0.13 & 0.06 & 0.028 & -0.19 & -0.01 \\
        Indirect Effect (ACME) & 0.21 & 0.10 & 0.036 & 0.03 & 0.41 \\
        Direct Effect (ADE) & 1.64 & 0.15 & 0.000 & 1.34 & 1.94 \\
        Proportion Mediated & 0.113& 0.05 & 0.051 & 0.22 & 0.04\\
        \midrule
        \textbf{Confidence} & & & & & \\
        AV Performance → Confidence & 2.03 & 0.18 & 0.000 & 1.68 & 2.39 \\
        Confidence → Trust & 0.52 & 0.04 & 0.000 & 0.44 & 0.61 \\
        Indirect Effect (ACME) & 0.87 & 0.12 & 0.000 & 0.65 & 1.14 \\
        Direct Effect (ADE) & 0.98 & 0.13 & 0.000 & 0.72 & 1.24 \\
        Proportion Mediated & 0.467 & 0.06 & 0.000 &0.36& 0.58\\
        \midrule
        \textbf{Loneliness} & & & & & \\
        AV Performance → Loneliness & -0.55 & 0.26 & 0.039 & -1.07 & -0.03 \\
        Loneliness → Trust & 0.09 & 0.04 & 0.008 & 0.03 & 0.17 \\
        Indirect Effect (ACME) & 0.01 & 0.02 & 0.498 & -0.02 & 0.09 \\
        Direct Effect (ADE) & 1.83 & 0.12 & 0.000 & 1.58 & 2.07 \\
        Proportion Mediated & 0.007& 0.01 &0.490 & -0.17 & 0.04\\
        \midrule
        \textbf{Anxiety} & & & & & \\
        AV Performance → Anxiety & -1.37 & 0.20 & 0.000 & -1.77 & -0.97 \\
        Anxiety → Trust & -0.24 & 0.05 & 0.000 & -0.33 & -0.15 \\
        Indirect Effect (ACME) & 0.31 & 0.08 & 0.000 & 0.18 & 0.50 \\
        Direct Effect (ADE) & 1.53 & 0.13 & 0.000 & 1.27 & 1.79 \\
        Proportion Mediated & 0.177& 0.04&0.000 &0.101 &0.26\\
        \bottomrule
    \end{tabular}
    \label{tab:mixed-regression}
\end{table}

\subsection{Physiological Responses and Their Relationship with Trust}
A two-way repeated-measures ANOVA was conducted to examine the effects of risk perception and AV performance on mean HR and mean SCR.

\textbf{HR:} Analysis of HR data revealed a significant main effect of AV performance ($F(1, 67) = 5.83$, $p = .019$, $\eta^2_p = .080$). Participants exhibited significantly elevated mean HR during exposure to the "with errors" condition compared to the "without errors" condition, potentially reflecting increased cognitive load or stress responses. Neither the main effect of risk perception ($F(2, 67) = 0.46$, $p = 0.636$, $\eta^2_p = 0.013$) nor the interaction between risk perception and AV performance ($F(2, 67) = 0.17$, $p = 0.847$, $\eta^2_p = 0.005$) reached statistical significance, indicating that prior risk information did not substantially influence cardiovascular responses during the experiment.

\textbf{SCR:} Analysis of the mean SCR yielded no statistically significant effects. There was no significant main effect of AV performance ($F(1, 67) = 0.51$, $p = 0.479$, $\eta^2_p = 0.007$) and risk perception ($F(2, 67) = 0.486$, $p = 0.618$, $\eta^2_p = 0.014$). Similarly, no significant interaction effect was observed ($F(2, 67) = 1.33$, $p = 0.271$, $\eta^2_p = 0.038$) in participants' electrodermal activity during the experiment. 

\begin{figure}[ht!] 
    \centering 
    \includegraphics[width=0.7\textwidth]{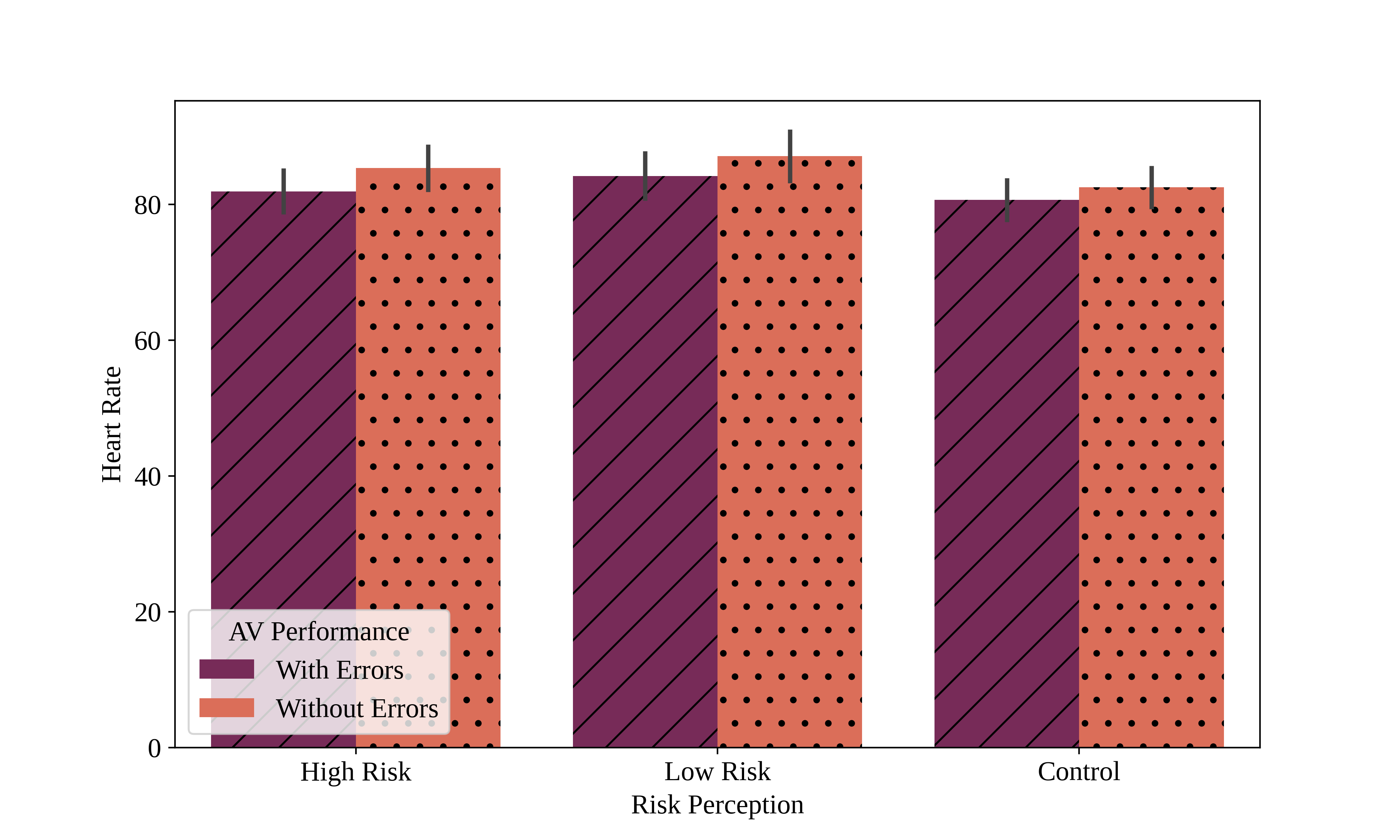} 
    \caption{Average Heart Rate across groups.} 
    \label{fig:heart_rate} 
\end{figure}



\subsubsection{Time-Series Analysis of Physiological Responses\\}

\begin{figure}
    \centering 
    \begin{subfigure}[b]{0.8\textwidth} 
        \centering 
        \includegraphics[width=\textwidth]{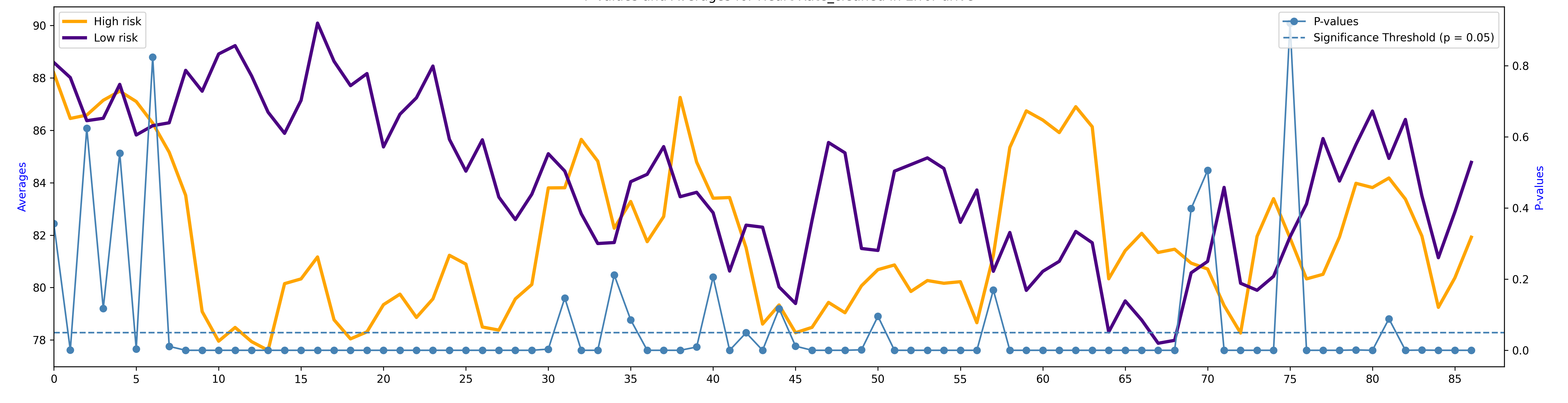} 
        \caption{HR dynamics in high-risk versus low-risk groups.} 
        \label{fig:hr_E} 
    \end{subfigure} 
    \vspace{0.7cm}
    \begin{subfigure}[b]{0.8\textwidth} 
        \centering 
        \includegraphics[width=\textwidth]{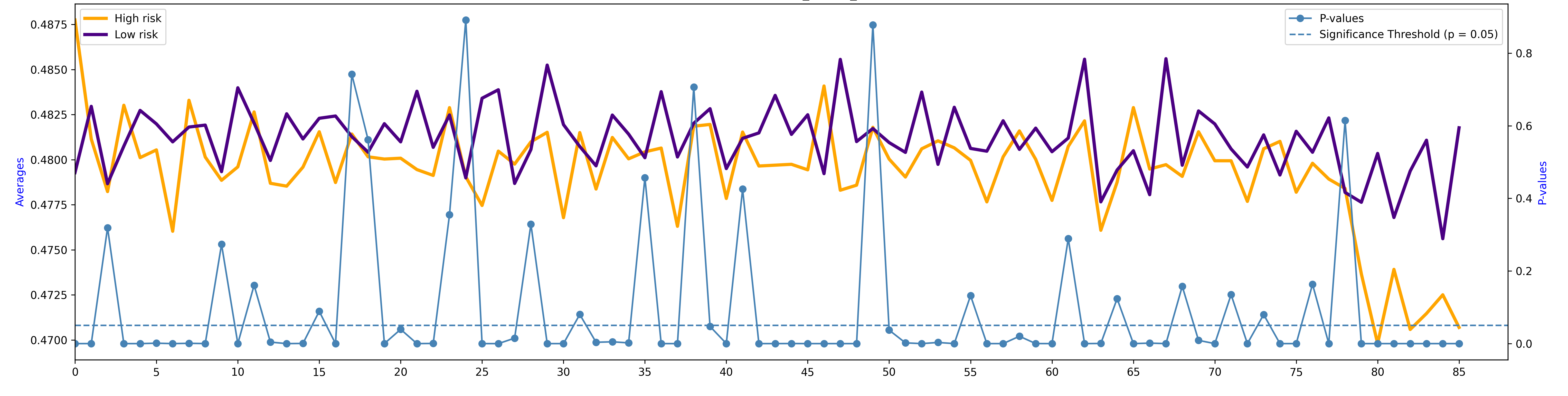} 
        \caption{GSR dynamics in high-risk versus low-risk groups.} 
        \label{fig:eda_E} 
    \end{subfigure} 
    \caption{Temporal dynamics and of physiological markers (HR and GSR) during error-prone driving scenarios, with corresponding statistical significance indicators.}
    \label{fig:hr_time}
\end{figure}

\begin{figure}
    \centering 
    \begin{subfigure}[b]{0.8\textwidth} 
        \centering 
        \includegraphics[width=\textwidth]{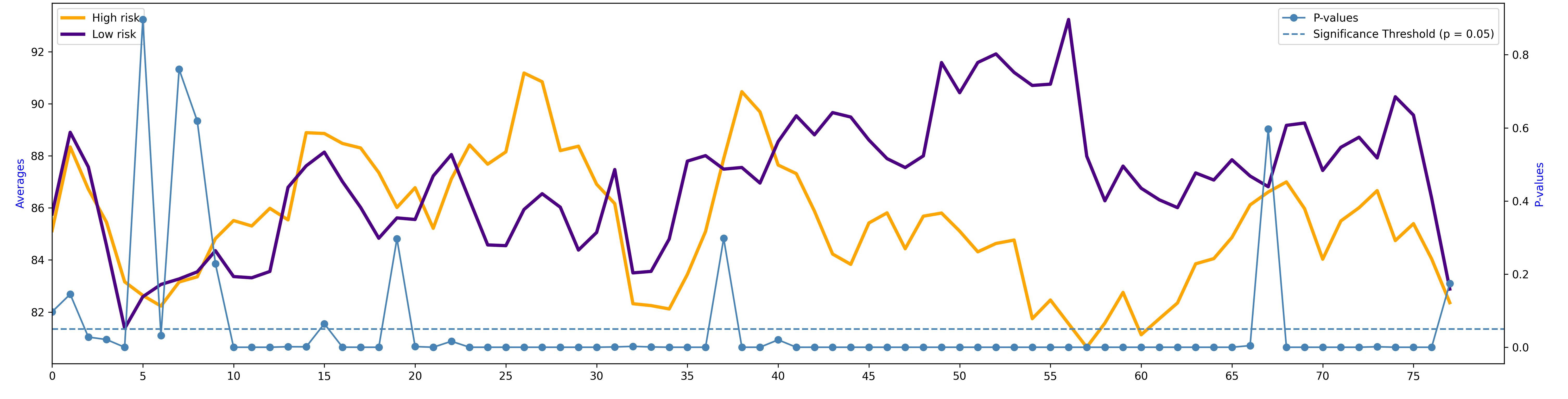}
        
        \caption{Heart rate dynamics in high-risk versus low-risk groups.} 
        \label{fig:hr_noE} 
    \end{subfigure} 
    \vspace{0.7cm}
    \begin{subfigure}[b]{0.8\textwidth} 
        \centering 
        \includegraphics[width=\textwidth]{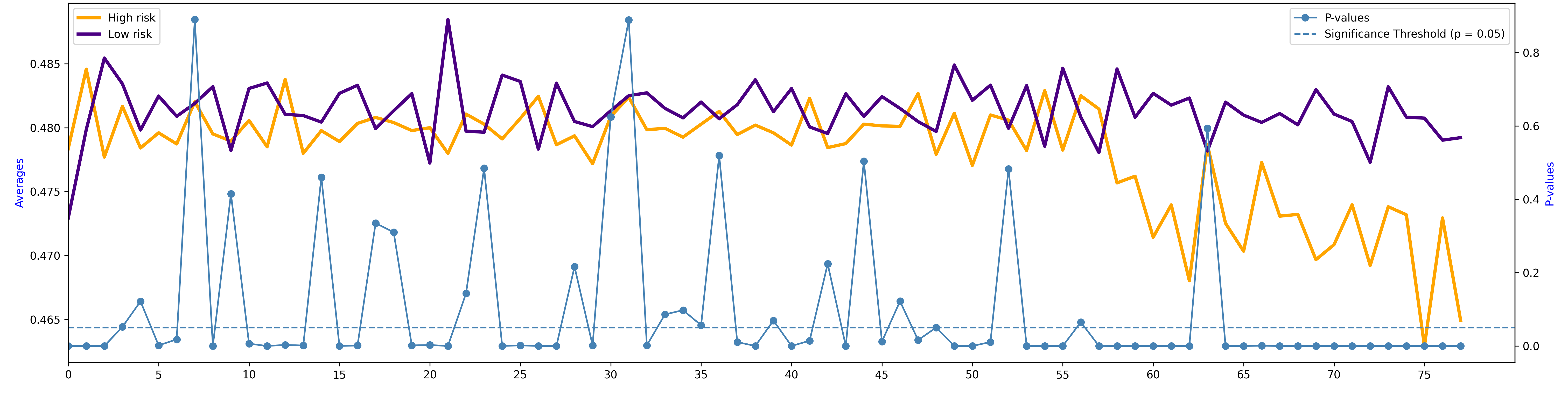}
        
        \caption{GSR dynamics in high-risk versus low-risk groups.} 
        \label{fig:eda_nE} 
    \end{subfigure} 
    \caption{Temporal dynamics and of physiological markers (HR and GSR) in high-risk versus low-risk groups during error-free driving scenarios, with corresponding statistical significance indicators.}
    \label{fig:eda_time}
\end{figure}

To examine the temporal dynamics of physiological responses, we segmented HR and SCR data into consecutive 10-second intervals and conducted a visual inspection of the resulting time-series data (see Figures \ref{fig:hr_time} and \ref{fig:eda_time}). The blue line in each figure represents statistical significance (p-values) between the high-risk and low-risk conditions at each time segment, with the dashed horizontal line indicating the significance threshold ($p = 0.05$). Note that the control condition was excluded from these temporal analyses to focus specifically on the contrast between high-risk and low-risk manipulations.
This fine-grained temporal analysis allowed us to identify subtle variations in physiological metrics that might be obscured when examining only aggregate data across entire trials, while also allowing us to correlate these physiological responses with specific events occurring within the video stimulus at specific time periods.

\textbf{Drive with With Errors:}
In the condition with system errors, participants exhibited distinctive physiological patterns indicating dynamic trust calibration throughout the AV experience. During the initial phase (segments 0-7), as the AV initiated startup and departed from the residential area, both risk groups displayed comparable HR measurements (see Figure \ref{fig:hr_E}). However, statistical analysis revealed significant differences ($p < 0.05$) at segments 5 and 7, where the low-risk group demonstrated lower HRs compared to the high-risk group. This pattern suggests that participants primed with high-risk information began with a heightened physiological state, potentially reflecting a more cautious baseline trust level and increased vigilance.
As the drive progressed to segments 8-10, the AV began experiencing navigation errors, triggering divergent physiological responses between groups. The high-risk condition showed a sharp decrease in HR, while the low-risk condition exhibited an increasing trend, resulting in statistically significant differences ($p < 0.05$) between groups. This pattern persisted as the AV hesitated at an intersection due to a closed lane, with low-risk participants maintaining elevated HR until they received system acknowledgment through a support call.
Between segments 20-30, HRs in both conditions gradually converged to approximately 84-86 BPM as vehicle assistance became closer. 

Two separate patterns were identified within the middle phase. The first period (segments 30-45) was characterized by alternating HR patterns corresponding to the AV's transitions between waiting states and movement, including an abrupt turn and stop that partially blocked the road before returning to a waiting state. During the second period (segments 45-70), the AV displayed incorrect information that was inconsistent with system and destination details. This inconsistency significantly elicited increased HR in the low-risk group, while the high-risk group maintained relatively stable, lower rates. This notable physiological response indicated a momentary breach of trust triggered by system errors, confirming that prior risk perception influences participants' sensitivity to performance failures.
A particularly notable event occurred at segment 56, when participants noticed the support vehicle approaching and observed road workers near the AV. At this point, low-risk participants experienced a decrease in HR, while the high-risk group showed an abrupt increase, resulting in significant differences between groups. This divergent response pattern suggested different trust recalibration mechanisms based on prior risk expectations.
In the final segments (70-86), the AV made an abrupt correction before resuming normal operation, triggering elevated HRs in both groups. However, the low-risk group exhibited significantly higher rates compared to the high-risk group. As the drive continued, HR began normalizing until the AV approached another hazard and demonstrated irregular patterns of behavior by moving and stopping, which again elevated HR in both groups. Throughout this final phase, the low-risk group maintained significantly higher HR than the high-risk group, suggesting persistent differences in physiological responses to system recovery after errors based on initial risk perception.

SCR measurements (see Figure \ref{fig:eda_E}) further substantiated these findings, with the low-risk condition demonstrating consistently higher SCR during error events. Statistical significance ($p < 0.05$) was observed across most segments, reinforcing the differential physiological responses between risk conditions.

\textbf{Drive with Without Errors:}
In error-free driving scenarios, we observed an unexpected trust development pattern with statistical significance ($p < 0.05$) across nearly all trial segments. During initial exposure (segments 0-20), both conditions exhibited similar HR fluctuations (84-86 BPM), likely reflecting adaptation to the AV navigating side roads. As the AV entered busier roads (segments 20-30), a notable divergence emerged with the high-risk condition showing elevated HR (88-91 BPM) compared to the low-risk condition (84-86 BPM). However, when navigating narrow roads with parked cars, the high-risk condition demonstrated a decline until re-entering major thoroughfares.
During segments 40-60, when the AV traveled consistently on a straight but busy road, an inverse relationship developed. The low-risk condition maintained elevated HRs while the high-risk condition showed a gradual decline, suggesting differential cognitive engagement patterns. This counterintuitive finding may indicate physiological adaptation in the high-risk group, as consistent error-free performance gradually re-calibrated the trust and eventually overcame initial risk perception.
Conversely, low-risk participants potentially experienced increased arousal when encountering traffic conditions that exceeded their initial expectations, despite the AV's flawless performance.
In the final phase (segments 60-75), as the AV executed lane changes and responded to traffic signals, the high-risk condition showed increased, then slightly fluctuating HRs (81-87 BPM). 
Concurrent SCR measurements showed that both high- and low-risk conditions exhibited consistent oscillatory patterns throughout most of the drive (see Figure \ref{fig:eda_nE}).  Statistical significance of these differences appeared as periodic spikes, indicating moments of heightened physiological distinction between the risk conditions. These periods largely coincided with specific driving events, such as approaching traffic lights or navigating turns within the neighborhood.  The divergence between high- and low-risk conditions became particularly pronounced in the later segments, as the AV completed final turns, executed lane changes, stoped at traffic signals, and provided rider notifications regarding arrival estimates and exit instructions.  These observed patterns suggested that participants' physiological responses might be influenced not only by immediate driving maneuvers, but also by the sequential complexity of automated driving tasks and system-passenger interactions.

\section{Discussions}
This study investigated the interplay of risk perception, AV performance (specifically, the presence or absence of automation errors), and emotional responses on drivers' trust in AVs. The findings provided support for the critical role of actual system performance in shaping trust, highlighted the mediating role of specific emotions, and suggested that while risk perception manipulations could influence initial trust, their effect was overshadowed by direct experience with the AV's behavior.

\subsection{Emotional Responses: A Four-Factor Structure}

The EFA of the 19 emotion items revealed a four-factor structure: Hostility, Confidence, Anxiety, and Loneliness. This structure aligned, in part, with prior research on emotions in human-automation interaction \citep{buck2018user, avetisian2022anticipated, jensen2020anticipated}. The prominence of Hostility as the factor explaining the most variance underscored the importance of negative emotional reactions in response to perceived risks and, especially, system errors. This finding was consistent with the literature emphasizing the disproportionate impact of negative events and emotions on trust formation and erosion \citep{choi2015investigating, lee2004trust}. The emergence of Confidence as a distinct factor reinforced the importance of positive affective experiences in building trust, a concept often less emphasized in traditional, cognitively focused trust models.


\subsection{AV Performance Dominates Risk Perception in Shaping Trust}

Addressing RQ1, the results clearly demonstrated that AV performance had a significant and substantial impact on trust. Participants' trust levels were significantly lower after experiencing the scenario with AV errors compared to the scenario with flawless performance. This finding supported the central role of direct experience and system reliability in shaping trust, as highlighted in the learned trust component of Hoff and Bashir's \citeyearpar{hoff2015trust} three-layered trust model.

Interestingly, while the risk perception manipulation did significantly influence initial learned trust, it did not have a significant direct effect on trust after experiencing the driving scenarios. Furthermore, there was no significant interaction between risk perception and AV performance on trust. This suggested that the immediate experience of the AV's behavior (reliable or unreliable) outweighed pre-existing beliefs about AV risks in determining trust levels during the interaction. This contrasted with some studies that emphasized the strong influence of risk perception \citep{gold2013take}, but it aligned with the idea that trust in automation was highly dynamic and responsive to real-time performance \citep{ayoub2021investigation}. It was possible that the relatively short duration of the interaction in this study prioritized immediate experience over pre-existing risk beliefs.

The finding that more experienced drivers exhibited slightly lower trust aligned with research suggesting that experienced drivers might be more critical of automation, potentially due to a greater sense of their own driving competence and a heightened awareness of potential inconsistencies between expected and actual AV behavior \citep{schoettle2014survey}.  The marginally significant positive effect of risk propensity on trust suggested that individuals with a greater tolerance for risk might be more willing to trust AVs, even in the face of uncertainty. Conversely, eagerness to adopt new technologies positively influenced trust, indicating that individuals more open to technological advancements were generally more receptive to AVs. These findings aligned with our previous findings \cite{Ayoub2021Modeling} and technology acceptance models \citep{venkatesh2012consumer}, which posited that openness to new technology strongly predicts trust and adoption.

\subsection{Mediation of Emotions in Trust Formation}

Addressing RQ2, the mediation analysis provided strong evidence that emotional responses mediated the relationship between AV performance and trust.  Specifically, Confidence emerged as the most significant mediator.  Error-free AV performance led to increased feelings of confidence, which, in turn, significantly increased trust in the AV. This finding highlights the important role of positive emotional experiences in fostering trust, supporting the growing recognition of affective trust as a complement to, and sometimes even a driver of, cognitive trust \citep{lee1992trust, madsen2000measuring, choi2015investigating}.

Hostility and anxiety also significantly mediated the relationship between AV performance and trust, but in a negative direction.  Error-free performance reduced feelings of hostility and anxiety, which, in turn, led to increased trust.  This aligned with research showing the detrimental impact of negative emotions on trust \citep{de2020towards}. However, the mediation effects for hostility and anxiety, while significant, were smaller than the effect for confidence. This suggests that while mitigating negative emotions is important, fostering positive emotions (confidence) may be a more powerful lever for building trust in AVs.

Loneliness, while significantly affected by AV performance (error-free performance reduced loneliness), did not significantly mediate the relationship between performance and trust. This suggests that while feelings of isolation may be relevant to the overall experience of interacting with an AV, they may not be directly linked to trust in the system's capabilities.

\subsection{Theoretical and Practical Implications}

This study contributes to the theoretical understanding of trust in automation in several ways. First, we found that emotional responses were key mediators in trust formation, while negative emotions might have a more complex impact on trust. The study clearly demonstrates the mediating role of specific emotions, particularly confidence, in the relationship between AV performance and trust. This provides a more nuanced understanding of how emotions influence trust with empirical support for integrating both cognitive (system performance, risk perception) and affective (emotional responses) factors in models of trust formation in AVs. Second, our findings also emphasize the dynamic and experience-based nature of trust, highlighting the dominance of real-time system performance over pre-existing risk perceptions in shaping trust during interaction.

These findings have several practical implications for the design and deployment of AVs. First, we should prioritize error-free performance as minimizing system errors is paramount for building and maintaining trust.  Even relatively minor errors can significantly erode trust. Second, we should design features that promote feelings of confidence (e.g., clear and transparent communication, smooth and predictable behavior), which are crucial for fostering trust. Third, while fostering confidence is key, designers should also address factors that contribute to negative emotions like hostility and anxiety (e.g., providing clear explanations for system actions, allowing for user control when appropriate).


\subsection{Limitations and Future Research}
Despite the valuable insights provided by this study, several limitations must be acknowledged. First, the sample consisted primarily of university students and young adults, limiting the findings' generalizability. Individuals with different demographic backgrounds, such as older adults or those with more extensive driving experience, may respond differently to AV performance and risk perception, and their emotional reactions may vary accordingly. Future research should aim to include a more diverse sample to enhance the external validity of the results.
Second, the study employed short-term exposure to AV performance through simulated driving scenarios. While this method effectively captures real-time emotional responses and trust, it does not fully represent how long-term interactions with AVs might influence trust calibration and emotional responses over time. Trust in automation is known to evolve through repeated interactions \citep{hoff2015trust}, and longitudinal studies would better capture the dynamic nature of trust development, particularly in response to both positive and negative system performance.
Finally, although the mediation analysis identified confidence as a significant mediator of the relationship between AV performance and trust, the role of other emotional components remained less clear. Future research could explore contextual factors or different methodological approaches (e.g., experimental manipulations of emotional states) to understand further how negative emotions may mediate trust over longer-term interactions with AVs or under more complex driving conditions.

\section{Conclusion}
In conclusion, this study demonstrates that automation errors significantly reduce trust in AVs, while risk perception plays a lesser role. Confidence emerged as a significant emotional mediator, suggesting that positive emotional experiences are crucial in the development of trust. These findings provide important insights into the emotional and cognitive factors that influence trust in automated systems and offer practical recommendations for improving user trust in AV technologies. Further research could explore how long-term interactions with AVs influence the development of trust and whether the impact of negative emotions on trust becomes more pronounced over extended periods of use.

\section{Acknowledgment}
This research was supported by the National Science Foundation.

\bibliography{main}
\newpage
\end{document}